\documentclass[9pt,twocolumn,twoside]{optica}
\setboolean{shortarticle}{false}
\setboolean{minireview}{false}

\usepackage{multirow}
\usepackage{amsmath,amssymb}
\usepackage{booktabs}
\usepackage{mathrsfs}
\usepackage{lipsum}
\usepackage{mathtools}
\usepackage{stackrel}
\usepackage{color, colortbl}
\usepackage{xcolor}

\usepackage{epsfig}

\newcommand{\ee}[1]{\text{e}^{#1}}
\newcommand{\expb}[1]{\exp\left\{ #1 \right\}}

\newcommand{\sst}[2]{{#1}_{\text{#2}}}

\newcommand{\breathe}[2]{\rule[-#1ex]{0cm}{#2ex}}

\newcommand{\ltwonorm}[1]{\left|\!\left|#1\right|\!\right|_2}

\newcommand\balphlist[1]{\newcounter{#1}
\begin{list}
  {{\bf \alph{#1})}}   
  {\usecounter{#1}\setlength{\rightmargin}{0cm}
   \itemsep 1ex\parsep 1ex}}
\newcommand\alphlist[1]{\newcounter{#1}
\begin{list}
  {{\alph{#1})}}   
  {\usecounter{#1}\setlength{\rightmargin}{0cm}
   \itemsep 1ex\parsep 1ex}}
\newcommand\bnumlist[1]{\newcounter{#1}
\begin{list}
  {{\bf \arabic{#1})}}   
  {\usecounter{#1}\setlength{\rightmargin}{0cm}
   \itemsep 1ex\parsep 1ex}}
\newcommand\numlist[1]{\newcounter{#1}
\begin{list}
  {{\arabic{#1})}}   
  {\usecounter{#1}\setlength{\rightmargin}{0cm}
   \itemsep 1ex\parsep 1ex}}
\newcommand\romlist[1]{\newcounter{#1}
\begin{list}
  {{(\roman{#1})}}   
  {\usecounter{#1}\setlength{\rightmargin}{0cm}
   \itemsep 1ex\parsep 1ex}}
\newcommand\bromlist[1]{\newcounter{#1}
\begin{list}
  {{\bf \roman{#1})}}   
  {\usecounter{#1}\setlength{\rightmargin}{0cm}
   \itemsep 1ex\parsep 1ex}}
\newcommand\problist{\newcounter{probc}
\begin{list}
  {{\bf \arabic{probc}.}}   
  {\usecounter{probc}\setlength{\rightmargin}{0cm}
   \itemsep 1ex\parsep 1ex}}
\newcommand\subproblist[1]{\newcounter{#1}
\begin{list}
  {{\bf \arabic{probc}.\alph{#1})}}   
  {\usecounter{#1}\setlength{\rightmargin}{0cm}
   \itemsep 1ex\parsep 1ex}}

\title{Learning to Synthesize: Robust Phase Retrieval at Low Photon Counts}

\author[1,*]{Mo Deng}
\author[2,$\dagger$]{Shuai Li}
\author[2,$\ddagger$]{Alexandre Goy}
\author[1]{Iksung Kang}
\author[2,3]{George Barbastathis}

\affil[1]{Department of Electrical Engineering and Computer Science, Massachusetts Institute of Technology, Cambridge, MA 02139, USA}
\affil[2]{Department of Mechanical Engineering, Massachusetts Institute of Technology, Cambridge, MA 02139, USA}
\affil[3]{Singapore-MIT Alliance for Research and Technology (SMART) Centre, Singapore 117543, Singapore}
\affil[$\dagger$]{Current address: SenseTime Research, 2550 N 1st Street, Suite 300, San Jose, CA 95131, USA}
\affil[$\ddagger$]{Current Address: Omnisens SA, Riond Bosson 3, 1110 Morges, VD, Switzerland}
\affil[*]{Corresponding author: modeng@mit.edu} 

\begin{abstract}
The quality of  inverse problem solutions obtained through deep learning [Barbastathis \textit{et al}, 2019] is limited by the nature of the priors learned from examples presented during the training phase. In the case of quantitative phase retrieval [Sinha \textit{et al} 2017, Goy \textit{et al} 2019], in particular, spatial frequencies that are underrepresented in the training database, most often at the high band, tend to be suppressed in the reconstruction.  {\it Ad hoc} solutions have been proposed, such as pre-amplifying the high spatial frequencies in the examples [Li \textit{et al}, 2018]; however, while that strategy improves resolution, it also leads to high-frequency artifacts as well as low-frequency distortions in the reconstructions. Here, we present a new approach that learns separately how to handle the two frequency bands, low and high; and also learns how to synthesize these two bands into the full-band reconstructions. We show that this ``learning to synthesize'' (LS) method yields phase reconstructions of high spatial resolution and artifact-free; and it is also resilient to high-noise conditions, e.g. in the case of very low photon flux. In addition to the problem of quantitative phase retrieval, the LS method is applicable, in principle, to any inverse problem where the forward operator treats different frequency bands unevenly, {\it i.e.} is ill-posed. 
\end{abstract}

\setboolean{displaycopyright}{true}

\begin{document}

\maketitle

\section{Introduction} \label{sec:intro}

The retrieval of phase from intensity is one of the most important and most challenging problems in Classical Optics. The utility of phase retrieval stems from the fact that it allows the shape of transparent objects, biological cells in particular, to be quantified in two and three spatial dimensions using visible light\ \cite{inv:marquet05,inv:popescu2006diffraction}. In the x-ray band, quantitative phase imaging is also useful because phase contrast in tissue is orders of magnitude higher than attenuation contrast\ \cite{inv:mayo03,inv:pfeiffer06}. The same argument can be made for identification of liquids\ \cite{inv:pan2014contrast} and semiconductor materials for integrated circuit characterization and inspection\ \cite{inv:holler17}.

There are two well-known challenges in phase retrieval. Firstly, for the phase of the optical field to be well-defined, the illumination needs to be temporally and spatially coherent to a fairly good approximation; this is especially difficult with x-rays. One way to relax this requirement is to acknowledge that it is usually the phase {\em delay through the object} that is of interest, and seldom the phase of the optical field itself. The former can be obtained even from partially coherent light\ \cite{petruccelli:2013} but requires the correlation function (mutual intensity) to be measured, which is often problematic because of low contrast. The second challenge is that, since only the intensity of a light beam is observable, the phase may only be inferred indirectly from intensity measurements. Computational approaches to this operation may be classified as interferometric/holographic\ \cite{inv:goodman1967digital,inv:creath1985phase}, where a reference beam is provided; and non-interferometric, or reference-less such as direct/iterative\ \cite{gerchberg:1972,inv:fienup1978reconstruction} and ptychographic\ \cite{zheng2013wide,tian2014multiplexed}, which are both nonlinear, and transport-based\ \cite{inv:teague1983deterministic,inv:Streibl84-TIE}, where the problem is linearized through a hydrodynamic approximation. Direct methods attempt to retrieve the phase from a single raw intensity image, whereas the transport and ptychographic methods implement axial and lateral scanning, respectively. What reference-less methods have in common is the need to obtain intensity measurements at some distance away from the conjugate plane of the object, {\it i.e.} with a small defocus. Direct measurement with defocus is the approach we take here. 

All computational phase retrieval approaches, interferometric and non-interferometric, involve the solution of a nonlinear and highly ill-posed inverse problem. For direct phase imaging, which is a nonlinear problem---see Section~\ref{sec:problem}.\ref{sec:pr}---the classical Gerchberg-Saxton-Fienup (GSF) algorithm\ \cite{gerchberg:1972,inv:fienup1978reconstruction,3di:fienup82} and its variants\ \cite{inv:bauschke2002phase} are widely used. The main idea is to start with a random estimate for the unknown phase distribution and then to iteratively update it until the modulus-squared of its Fourier (or Fresnel) transform matches the observed intensity. For well-behaved phase fields, the iteration usually converges to the correct phase\ \cite{inv:Gerchberg86lock,inv:fienup1986phase}. Alternatively, the Wiener-Tikhonov functional minimization approach, described in Section~\ref{sec:problem}.\ref{sec:inverse}, exploits prior knowledge about the class of phase objects being imaged to combat noise artifacts. A modern and popular implementation is compressive imaging\ \cite{inv:candes-tao05,inv:candes2006robust,inv:donoho2006compressed,inv:candes-romberg-tao06b,book:eldar12} that utilizes sparsity priors, valid when the object is expressed in the appropriate set of basis functions. If the sparsifying set is unknown, it can be learned from examples according to the $k$-SVD method or other dictionary approaches\  \cite{elad2006image,inv:aharon2006k,inv:rubinstein2010dictionaries,inv:bao16}. 

In 2010, \cite{gregor2010learning} proposed a deep neural network in a recurrent scheme to learn the prior from examples, as an alternative to the dictionary approach. Subsequently, generator-discriminator competition\ \cite{nn:gan} was adopted as a more secure means of learning the statistical relationship between valid objects and their forward models; and the recursion was unfolded into a cascade for better numerical stability\ \cite{mardani:2017}. In these schemes, the physical model of the measurement is taken explicitly into account as a projection operator applied to the reconstruction estimate repeatedly at each recursion or cascade stage. This generalization of dictionaries to deep learning has been successful in a number of linear inverse problems, most notably superresolution\ \cite{yang2014single,inv:dong14-super-res,inv:dong15-super-res} and tomography\ \cite{inv:Jin2017,inv:McCann2017}. 

Recently, deep learning regression has been investigated for application to nonlinear inverse problems, in particular phase retrieval: direct\ \cite{inv:sinha17-PhENN,inv:PhENN-spectral-premod,inv:goy2018low}, holographic\ \cite{inv:rivenson18-dldh,wu2018extended}, and ptychographic\ \cite{nguyen2018deep,xue2019reliable}. The idea, described briefly in Section~\ref{sec:problem}.\ref{sec:inverse}, is to train a deep neural network (DNN) in supervised mode from examples of phase objects and their intensity images so that, after training, given an intensity image as input, the DNN outputs an estimate of the phase object. In this case, the physical model is either learned implicitly together with the prior from the examples\ \cite{inv:sinha17-PhENN,inv:PhENN-spectral-premod}; or incorporated as a pre-processor (``Approximant'')\ \cite{inv:goy2018low,inv:rivenson18-dldh,wu2018extended}. An interesting alternative method for the inverse problem, also nonlinear, of reconstructing the three-dimensional (3D) refractive index distribution from intensity projections, is to define the DNN architecture according to the strong scattering model and store the refractive index values as weights of the DNN after training\ \cite{inv:kamilov15-learn-tomo}. This ``index-storing'' DNN itself was subsequently used as Approximant to a traditional DNN for improving the estimates in 3D distributions with exceptionally small and high-contrast features or when the range of available angles of projection is severely limited\ \cite{inv:goy19-3Dtomo}. Extensive reviews of deep learning use for inverse problems can be found in \cite{inv:McCann2017,inv:Jo18-review,inv:barbastathis19-review}.

Here, we propose a new DNN-based computational architecture for phase retrieval with the unique feature of processing low and high spatial frequency bands as separate channels with two corresponding DNNs trained from an original object database and a high-pass filtered version of the database, respectively. Subsequently, the outputs of the two channels are recombined using a third DNN also specifically trained for this task. The motivation for this new approach is an earlier observation\ \cite{inv:PhENN-spectral-premod} that nonlinearities in DNN training and execution algorithms tend to amplify imbalances in the spatial frequency content of the training database and in the way different spatial frequencies are treated as they propagate through the physical optical system; this amplified imbalance typically results in the lower spatial frequencies becoming dominant and ultimately limiting resolution of fine spatial features in the reconstructions. A more detailed overview of this phenomenon can be found in Section~\ref{sec:problem}.\ref{sec:train}. Because the essential feature of our new proposed technique is the synthesis of the two spatial bands through a trained DNN, we refer to it as ``learning to synthesize'' (LS). 

Splitting the spatial frequency content into several bands and processing the bands separately has a long history in signal processing\ \cite{inv:daubechies88-wavelets,books:daubechies92,inv:coifman95,book:strang96,inv:chan03,inv:daubechies04,inv:figueiredo03,book:mallat08}. For image reconstruction, dual band processing has been used in fluorescence microscopy\ \cite{inv:lim08-HiLo,inv:mertz11,inv:Bhattacharya12-3Dhilo} and phase retrieval\ \cite{inv:zhu2014low}. However, these cases, unlike ours, required structured illumination. In the context of learning-based inversion, a dual channel method has been tried for superresolution\ \cite{pan2018learning} (to be understood as upsampling) albeit the two processed channels were combined as a simple convex sum to form the final image. By contrast, the LS method presented here uses a {\em learned} nonlinear filter, implemented as a third DNN trained to optimally recombine the two channels according to the spectral properties of the class of objects that the training database represents. 

In addition to requiring a single raw image to retrieve the phase through a learned recombination of the spectral channels, the LS method presented here has the desirable property of resilience to noise, especially in the case of weak photon flux down to a single photon per pixel. We achieved that by using an Approximant filter\ \cite{inv:goy2018low} to pre-process the raw image before submitting it to the two spectral channels. The Approximant produces an inverse estimate that expressly uses the physical model (a single iteration of the GSF algorithm in \cite{inv:goy2018low} and here). For very noisy inputs, the Approximant is of very poor quality; yet, if the subsequent learning architecture is trained with this low-quality estimate as input, the final reconstruction results are significantly improved. The LS method with Approximant, as presented here, drastically improves over \cite{inv:goy2018low}, especially in the reconstruction of fine detail, as \cite{inv:goy2018low} did not use separate spectral channels to rebalance frequency content. 

The detailed implementation of the LS method is described in Section~\ref{sec:methods}, where we also show how to optionally include the Approximant as initial estimate for input to the LS scheme. Results from experiments are presented in Section~\ref{sec:results} with detailed characterization of the LS method's behavior under different noise conditions. Conclusions and suggestions for future work are in Section~\ref{sec:conclusions}.

\section{Problem formulation}
\label{sec:problem}

\subsection{Phase retrieval \label{sec:pr}}

Let
\[
\sst{\psi}{obj}(x,y)=t(x,y)\ee{if(x,y)}
\]
denote the complex transmittance of an optically thin object of modulus response $t(x,y)$ and phase response $f(x,y)$, and let $\sst{\psi}{inc}(x,y)$ denote the coherent incident field of wavelength $\lambda$ on the object plane. The noiseless intensity measurement $g_0(x,y)$ (also referred to as noiseless raw image) is carried out on the detector plane located at a distance $z$ away from the object plane, and can be written as 
\begin{equation}
g_0(x,y)=\left|\:\mathbf{F}_{z}\!\left[\sst{\psi}{inc}(x,y) \sst{\psi}{obj}(x,y)\right]\right|^2\equiv H_0f(x,y), 
\label{pr-forward-model}
\end{equation}
where $\mathbf{F}_{z}[\cdot]$ denotes the Fresnel (paraxial) propagation operator for distance $z$, {\it i.e.} the convolution
\begin{equation}
\mathbf{F}_{z}[\psi]=\psi(x,y)\star \frac{\ee{i2\pi z/\lambda}}{i\lambda z}\:\expb{i\pi\frac{x^2+y^2}{\lambda z}}; 
\label{eq:Fresnel}
\end{equation}
and $H_0$ is the (nonlinear) noiseless forward operator. Alternatively, $\mathbf{F}_z$ may be expressed in the spatial frequency domain $(\nu_x,\nu_y)$ as
\begin{equation}
\mathbf{F}_{z}[\psi] = {\cal F}^{-1}\left\{\breathe{2}{5}
{\cal F}\left\{\psi\right\} \: \expb{-i\pi\lambda z\left(\nu_x^2+\nu_y^2\right)}
\right\},
\label{eq:Fresn-spfr}
\end{equation}
where ${\cal F}$ denotes the 2D (spatial) Fourier transform operator and ${\cal F}^{-1}$ its inverse. 

We are interested in weakly absorbing objects, {\it i.e.} we assume $t(x,y)\approx1$. In all the experiments described here, the illumination is also a normally incident plane wave $\sst{\psi}{inc}(x,y)=1$. Therefore, to a good approximation, we may write 
\begin{equation}
g_0(x,y)=H_0f(x,y)=\left|\:\mathbf{F}_{z}\!\left[\ee{if(x,y)}\right]\right|^2.
\end{equation}
This is what we refer to as the direct phase retrieval problem, which Gerchberg-Saxton and related algorithms solve iteratively \cite{gerchberg:1972,3di:fienup82}. 

In practice, the measurement is subject to Poisson statistics due to the quantum nature of light; and to Gaussian thermal noise added by the photoelectric conversion process. We express the noisy measurement as 
\begin{equation}
g(x,y)=\mathscr{P}\left\{p\:\frac{H_0f(x,y)}{\left<H_0f\right>}\right\}+\mathscr{N}\equiv Hf,
\label{eq:noisy-forward-model}
\end{equation}
where $\mathscr{P}\{\theta\}$ denotes a Poisson random variable with mean $\theta$ and $\mathscr{N}$ a Gaussian random variable with zero mean and variance $\sigma^2$. The photon flux in photons per pixel per frame is denoted as $p$ and the spatial average $\left<H_0f\right>=\left<g_0\right>$ of the noiseless raw image in the denominator is necessary as normalization factor. The noisy forward operator is $H$ and the purpose of phase retrieval is to invert $H$ so as to recover $f$ as accurately as possible, despite the nonlinearity and randomness present in the measurements.

\subsection{Solution of the inverse problem}\label{sec:inverse}

The Wiener-Tikhonov approach to solving inverse problems of the form $g=Hf$ is to obtain the estimate $\hat{f}$ of the inverse as
\begin{equation}
\hat{f}=\stackrel[f]{}{\text{argmin}}\left\{\breathe{1}{3} D\left(H_0f,g\right)+\Phi(f)\right\}.
\label{eq:wt}
\end{equation}
Here, $D(H_0f,g)$ is the fitness (or data-fidelity) term and $D$ is a distance operator. Under the assumption of solely additive Gaussian noise, it is appropriate to use the $L^2$ norm
\begin{align}
D\left(H_0f,g\right)  & =\ltwonorm{H_0f-g}^2 \nonumber \\
& = \sum_{x,y}\left[\breathe{1}{3}H_0f(x,y)-g(x,y)\right]^2.
\end{align}
Somewhat less rigorously, but conveniently for mathematical manipulations, the same $L^2$ metric is used for the fitness term even for measurements strongly subject to non-Gaussian statistics, as we consider here. The topic of appropriate fitness metrics is beyond the scope of this paper; in any case, when machine learning is used to approximate (\ref{eq:wt}), the dilemma of choosing a metric shifts to  the loss function for training a deep neural network. We will address this latter problem in some detail in Section~\ref{sec:methods}.\ref{sub:LS}. 

The second term $\Phi(f)$ in (\ref{eq:wt}) is the regularizer, or prior knowledge term. Its purpose is to compete with the fitness term in the minimization so as to mitigate ill-posedness in the solution. That is, the regularizer penalizes solutions that are promoted by the noise in the forward problem, as in (\ref{eq:noisy-forward-model}) for example, but does not meet general criteria known {\it a priori} for valid objects. The prior may be defined explicitly, e.g. as a minimum energy\ \cite{inv:tikhonov63a,inv:tikhonov63b} or sparsity\ \cite{inv:candes-tao05,inv:candes2006robust,inv:donoho2006compressed,inv:candes-romberg-tao06b,book:eldar12} criterion; or learned from examples as a dictionary\ \cite{elad2006image,inv:aharon2006k,inv:rubinstein2010dictionaries, inv:bao16} or through a deep learning scheme\ \cite{gregor2010learning,mardani:2017,yang2014single,inv:dong14-super-res,inv:dong15-super-res,inv:McCann2017,inv:Jin2017,inv:sinha17-PhENN,inv:PhENN-spectral-premod,inv:goy2018low,inv:rivenson18-dldh,wu2018extended,nguyen2018deep,xue2019reliable}.

Here, as in earlier works on direct phase retrieval\ \cite{inv:sinha17-PhENN,inv:PhENN-spectral-premod,inv:goy2018low,inv:rivenson18-dldh,wu2018extended,nguyen2018deep,xue2019reliable}, and due to the nonlinearity of the forward model, we adopt the End-to-End and Approximant methods. These we denote as
\begin{align}
\text{End-to-End:} \quad & \hat{f}=\text{DNN}(g); \quad \text{and} \label{eq:dnn-e-e} \\
\text{Approximant:} \quad & \hat{f}=\text{DNN}(\hat{f}^*), \label{eq:dnn-a}
\end{align}
where $\text{DNN}(\cdot)$ is the output of a deep neural network. In the End-to-End approach, the burden is on the DNN to learn from examples both the forward operator $H$ and the prior $\Phi$ so as to execute, in one shot, an approximation to the ideal solution (\ref{eq:wt}). Training takes place in supervised mode, with known pairs of phase objects $f$ and their raw intensity images $g$ generated on a phase spatial light modulator (SLM) and measured on a digital camera, respectively. Note that training is generally very slow, taking several hours or days if a few thousand examples are used. However, after training is complete, the execution of (\ref{eq:dnn-e-e}) or (\ref{eq:dnn-a}) is very fast as it only requires forward (non-iterative) computations. This is one significant advantage over the standard way of minimizing the Wiener-Tikhonov functional (\ref{eq:wt}) iteratively for each image. 

When the inverse problem becomes severely ill-posed or the noise is extremely strong, the learning burden on the DNN becomes too high; then, generally, better results are obtained by training the DNN to receive as input the Approximant $\hat{f}^*$ instead of the raw measurement $g$ directly. The Approximant is obtained through an approximate inversion of the forward operator; for example, in \cite{inv:rivenson18-dldh} it was implemented as a digital holographic backpropagation algorithm, whereas in \cite{inv:goy2018low} it was the outcome of a single iteration of the Gerchberg-Saxton algorithm \cite{gerchberg:1972}. While these Approximants $\hat{f}^*$ generally do not look very good, especially in highly noisy situations \cite{inv:goy2018low}, through training the DNN is able to learn a better association of $\hat{f}^*$ with its corresponding true object $f$ than it can learn with the noisy raw measurement $g$. 

\subsection{Spectral properties of training \label{sec:train}}

The design of deep neural networks is an active field of research and a comprehensive review of methods and caveats is well beyond the scope of this paper. We refer the reader to \cite{inv:McCann2017,inv:Jo18-review,inv:barbastathis19-review} for more extensive background and references. Here, we discuss the influence on the quality of training of the spatial power spectral density of the database where example are drawn from.  

\vspace{0.5ex}

In both End-to-End and Approximant methods (\ref{eq:dnn-e-e}-\ref{eq:dnn-a}) the examples determine the object class prior to be learned by the DNN. For instance, if we train on examples drawn from the ImageNet database \cite{nn:russakovsky15-ImangeNet} of natural objects, the prior learned is weaker than if we train on the MNIST \cite{nn:lecun2010mnist} database of hand-written digits or a database of integrated circuit (IC) segments because both latter shapes are more restricted than natural images and with stronger spatial correlations across. Recent work \cite{inv:goy2018low,inv:IDiffNet} has shown that, unsurprisingly, training with stronger priors results in more robust reconstructions as long as the test objects are drawn from the same class. 

In \cite{inv:PhENN-spectral-premod}, we addressed the influence of the spatial power spectral density (PSD) $S(\nu_x,\nu_y)$ of the example database on the quality of training. It is well known\ \cite{inv:olshausen1996,inv:olshausen96b,inv:van1996modelling,inv:olshausen97,inv:lewicki99,inv:lewicki00} that two dimensional (2D) images of natural objects, such as those contained in ImageNet\cite{nn:russakovsky15-ImangeNet}, follow the inverse quadratic PSD law
\begin{equation}
S\left(\nu_x,\nu_y\right)=\frac{1}{\nu_x^2+\nu_y^2}.
\label{eq:psd}
\end{equation}
Other types of object classes of practical interest exhibit similar power-law decay, perhaps with slightly different exponents. This means that, if a neural network is trained on such an object class, higher spatial frequencies are presented less frequently to the DNN during the training stage. At face value, this is as it should be, since the relative popularity of different spatial frequencies in the database is precisely one of the priors that the DNN ideally should learn. 

\vspace{0.5ex}

This understanding needs to be modified in the context of inverse problems because the representation of high spatial frequencies in the raw images is also uneven---typically, to the high spatial frequencies' disadvantage. In the specific case of phase retrieval, higher spatial frequencies within the spatial bandwidth (as determined by the numerical aperture NA) have uniform transmission modulus but are more severely scrambled by the chirped oscillations of the transfer function (\ref{eq:Fresn-spfr}). Thus, higher spatial frequencies suffer a double penalty\ \cite{inv:PhENN-spectral-premod}: their recovery becomes more sensitive to noise due to the scrambling; and they are less popular due to the inverse-square (or similar) PSD law so they are presented less frequently to the DNN's training process. Moreover, since the DNN itself and its training routine are both highly nonlinear, there is an acute risk that any unevenness in the treatment of different spatial frequency bands may be amplified in the final result, eventually causing the lower frequencies to dominate. 
\vspace{0.5ex}
Ref. \cite{inv:PhENN-spectral-premod} attributed the inability of the Phase Extraction Neural Network (PhENN) \cite{inv:sinha17-PhENN} to resolve spatial features well within its admitted spatial bandwidth to this unequal treatment of spatial frequencies. They showed that PhENN's resolution is approximately doubled by pre-filtering the training examples as to flatten their PSD. That is, during the training, each example $f(x,y)$ from the database was replaced with its filtered version
\begin{equation}
\sst{f}{p}(x,y):={\cal F}^{-1}\left\{\breathe{1}{3}
{\cal F}\left\{f(x,y)\right\}\times C\left(\nu_x,\nu_y\right)
\right\}.
\label{eq:hpf}
\end{equation}
The transfer function was defined as the high-pass filter
 \begin{equation}
    C\left(\nu_x,\nu_y\right)=\sqrt{\nu_x^2+\nu_y^2}
    \label{eq:tf-half}
    \end{equation}
exactly compensating for the inverse-quadratic dependence (\ref{eq:psd}) and flattening the spectrum. Raw images for training were correspondingly filtered as
\[
g_p(x,y)=H\sst{f}{p}(x,y),
\] 
whereas, during the test, the un-filtered measurements ({\it i.e.}, as received from the camera) were used to obtain the reconstructions. Unfortunately, with this implementation amplification of high spatial frequency features, especially of artifacts caused even by weak noise, was also evident in the reconstructions. This is not surprising, since technically (\ref{eq:hpf}) trades off violating the prior in return for finer spatial resolution. The LS approach that we describe in the next section is meant to fix this problem. 

\section{Methodology of Learning to Synthesize}\label{sec:methods}

\subsection{Description of the LS-DNN system} \label{sub:LS}

In \cite{deng2018learning}, we proposed a two-step approach to tackle the difficulty of restoring high frequency components without introducing significant artifacts and distortions. Here we describe the two steps for training and execution in detail, as well as the design of the DNNs involved in the LS system. For unity in notation, we denote the input to the entire LS system as $\xi(x,y)$, to be understood as
\begin{equation}
\xi(x,y)=\left\{
\begin{array}{rl}
g(x,y) & \text{in the End-to-End scheme, and} \\
\hat{f}^*(x,y) & \text{in the Approximant scheme.}
\end{array}
\right.
\end{equation}
We will discuss the Approximant implementation in more detail in section~\ref{sec:physics-informed} below. 

\begin{figure}[hbt!]
\centering
\includegraphics[width=0.45\textwidth]{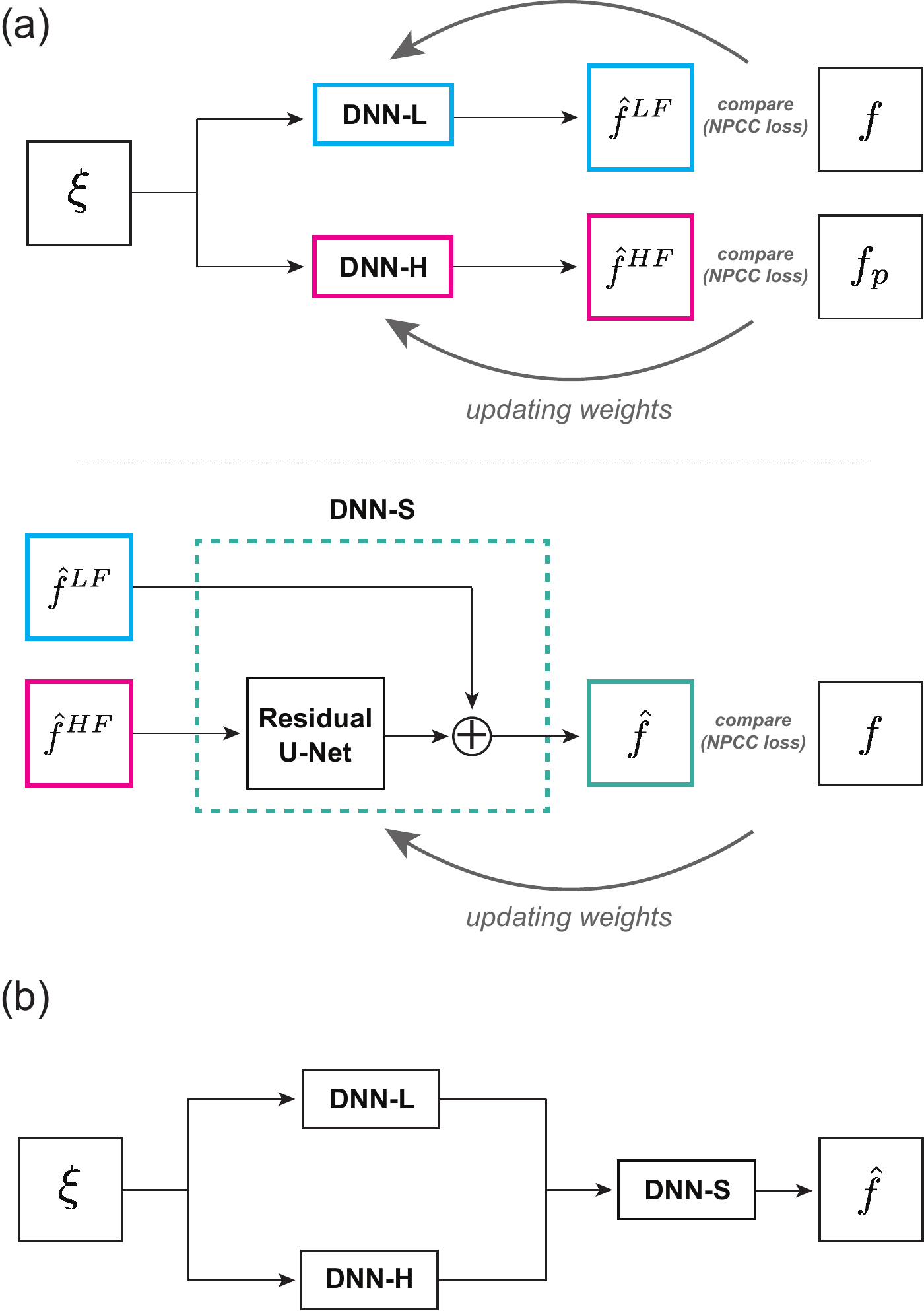}
\caption{LS-DNN schematic. (a) Training stage, (b) Test stage.}
\label{schematic}
\end{figure}

The two training steps are shown in block-diagram form in Figure~\ref{schematic}. The first step consists of training two separate DNNs in parallel, as follows: \\
\begin{itemize}
    \item DNN-L is trained to match unfiltered patterns $\xi^{(n)}(x,y)$ at its input with the corresponding unfiltered example phase patterns $f^{(n)}(x,y)$ as ground truth at its output (the superscript $n$ enumerates the examples). 
    \item DNN-H is trained to match unfiltered patterns $\xi^{(n)}(x,y)$ at its input to corresponding spectrally filtered versions $\sst{f}{p}^{(n)}(x,y)$ of the ground truth examples $f^{(n)}(x,y)$ at its output. The filter's transfer function is chosen more generally than (\ref{eq:tf-half}) as
   \begin{equation}
    C\left(\nu_x,\nu_y\right)=\left(\nu_x^2+\nu_y^2\right)^q.
    \label{eq:tfq}
    \end{equation}
    We have investigated implementations with $q$ spanning a broad range, as we discuss in more detail below. 
\end{itemize}

The output of DNN-L for a general test input $\xi(x,y)$ is denoted as $\hat{f}^{\text{LF}}(x,y)$. Assuming similar training conditions, $\hat{f}^{\text{LF}}$ matches the output of PhENN as presented in \cite{inv:sinha17-PhENN} in the End-to-End scheme or \cite{inv:goy2018low} in the Approximant scheme; that is, $\hat{f}^{\text{LF}}$ is expected to be fairly accurate at low spatial frequencies but missing fine detail. 

The output of DNN-H is denoted as $\hat{f}^{\text{HF}}(x,y)$. Note that \cite{inv:PhENN-spectral-premod} required spatial pre-filtering the raw inputs $g$; here, we do not spatially pre-filter the input $\xi$ ({\it i.e.}, $g$ or $\hat{f}^*$ according to whether the End-to-End or Approximant scheme is used). We instead train DNN-H to produce the filtered output based on an unfiltered input. This leads to better generalization, because DNN-H is trained on the broadest set of possible images (whereas the training in \cite{inv:PhENN-spectral-premod} was on high-frequency containing images only). Moreover, using unfiltered inputs to DNN-H allows for the training process to be parallelized for better efficiency. 

Depending on the value of $q$, the PSD of the patterns training DNN-H will be flat or almost flat. The output of DNN-H $\hat{f}^{\text{HF}}(x,y)$ is expected to have better fidelity at fine spatial features of the phase objects. However, spectral flattening may also generate artifacts due to over-learning the high spatial frequencies. Therefore, $\hat{f}^{\text{HF}}$ looks rather like a high-pass filtered version of the true object $f$, which we found to be more beneficial for subsequent use in the LS scheme. 

The second training step consists of combining the two partially accurate reconstructions $\hat{f}^{\text{LF}}$ and $\hat{f}^{\text{HF}}$ into a final estimate $\hat{f}(x,y)$ with uniform quality at all spatial frequencies, low and high up to the passband. To this end, we train the synthesizer DNN-S to receive $\hat{f}^{\text{LF}}$ and $\hat{f}^{\text{HF}}$ as inputs, and use the un-filtered examples $f$ as the output. To avoid any further damage to the high-spatial frequency content in $\hat{f}^{\text{HF}}$, we bypass  $\hat{f}^{\text{HF}}$ and present it intact to the last layer of DNN-S. By operating on $\hat{f}^{\text{LF}}$ alone, DNN-S learns how to treat the low frequency reconstruction so as to compensate for artifacts at all bands. The use of the synthesizer DNN-S also makes our results less sensitive to the choice of power $q$ in the transfer function (\ref{eq:tfq}). We found that $q \in [0.3,0.7]$ can produce reconstructions of approximately equal quality, as will be presented in section \ref{sec:results}. 

After DNN-L, DNN-H and DNN-S have been trained, they are combined in the LS system and operated as shown in Figure~\ref{schematic}(b). The input $\xi(x,y)$ is passed to DNN-L and DNN-H in parallel fashion, and the respective outputs  $\hat{f}^{\text{LF}}(x,y)$ and $\hat{f}^{\text{HF}}(x,y)$ are passed to DNN-S, which produces the final estimate $\hat{f}(x,y)$. It is worth noting that it is {\em not} valid to lump the three networks in Figure~\ref{schematic}(b) into a single network, due to their separate training schemes described above. 


There is a wide variety of DNN structures one may choose to implement DNN-L, H and S. In this work, we use for DNN-L the same architecture in \cite{inv:goy2018low}, a residual U-net architecture with skip connections \cite{RFB15a}. For simplicity, DNN-H and DNN-S, are also chosen to be structures similar to DNN-L. The details of the implementations are in the Supplementary Material. We made these choices to enable specific and fair comparisons with the earlier works; alternative architectures are certainly possible within the LS scheme, though we judged a full exploration to be outside the scope of the present paper.  

Training a neural network is typically implemented as a stochastic optimization procedure \cite{bottou2010large, Kingma2015AdamAM} where the neural network's internal coefficients (weights) are adjusted so as to minimize a metric of the distance between the actual output of the neural network and its desired output to a given input (training example). This distance is called training loss function (TLF). In the context of training to solve an inverse problem, the TLF is defined as
\begin{equation}
{\cal L}=\sum_n D(\hat{f}^{(n)}, f^{(n)}),
\end{equation}
where the superscript $n$ is again used to enumerate the examples, and the dilemma of choosing the appropriate metric operator $D$ emerges.

It is generally accepted \cite{nn:hinton-weight-decay87,nn:bishop06,inv:perceptual-loss,nn:goodfellow-dl-book,inv:ledig17,inv:sinha17-PhENN} that the $L^2$ metric (also referred to as mean square error, MSE) is a poor choice that does not generalize well; {\it i.e.}, deep neural networks trained with MSE do not perform well when presented with examples outside their training set. For image classification tasks, and in some early work on phase retrieval \cite{inv:sinha17-PhENN}, the $L^1$ metric (mean absolute error, MAE) was used instead. In direct analogy with compressive sensing, the $L^1$ metric promotes sparsity {\em in the internal connectivity} of the neural network, and that leads to better generalization. However, \cite{inv:IDiffNet} found that in highly ill-posed problems this benefit is eclipsed by the inability of MAE and pixel-wise metrics more generally to learn spatial structure priors about the object class that are crucial for regularization. 

In this paper, we train DNN-L, H and S using the Negative Pearson Correlation Coefficient (NPCC)\ \cite{inv:pearson1896,inv:rodgers88-pearson13,inv:IDiffNet} as the TLF. This is defined as
\begin{equation}
\sst{D}{NPCC}(a,b)\equiv
\frac {(-1)\times\displaystyle{\sum_{x,y}}\left(\breathe{1}{3} a(x,y)-\left<a\right>\right)\:
\left(\breathe{1}{3} b(x,y)-\left<b\right>\right)}
{\sqrt{\displaystyle{\sum_{x,y}}\left(\breathe{1}{3} a(x,y)-\left<a\right>\right)^2}\:
\sqrt{\displaystyle{\sum_{x,y}}\left(\breathe{1}{3} b(x,y)-\left<b\right>\right)^2}},
\label{eq:npcc}
\end{equation}
where $\left<a\right>$ and $\left<b\right>$ are the spatial averages of the generic functions $a(x,y)$, $b(x,y)$. If $a$ and $b$ are uncorrelated, the expected value of $\sst{D}{NPCC}(a,b)$ is zero, whereas if they are identical then $\sst{D}{NPCC}(a,b)=-1$. Thus, training the neural network minimizes the TLF toward ${\cal L}\approx -N$, where $N$ is the number of training examples.

The NPCC has been shown \cite{li2019analysis} to be more effective in recovering fine features than conventional loss functions such as the Mean Square Error (MSE), Mean Absolute Error (MAE) and the Structural Similarity (SSIM) index \cite{wang2003multiscale,inv:wang04-SSIM}. However, the NPCC has the disadvantage that it is invariant to affine transformations to its arguments, {\it i.e.} 
\begin{equation}
\sst{D}{NPCC}(a,b)=
\sst{D}{NPCC}(\alpha_1 a+\alpha_2,\beta_1 b+\beta_2)
\end{equation}
for arbitrary real numbers $\alpha_1$, $\alpha_2$, $\beta_1$, $\beta_2$. For {\em quantitative} phase retrieval, where the scale of phase difference matters, the affine ambiguity is resolved with a histogram equalization step after inversion \cite{inv:PhENN-spectral-premod}.

\subsection{Computation of the Approximant}\label{sec:physics-informed}

It has been shown that, even under extreme noise conditions, just a single iteration of the Gerchberg-Saxton (GS) algorithm suffices as Approximant in scheme (\ref{eq:dnn-a}) for phase retrieval\ \cite{inv:goy2018low}. We elected to use the same approach here for the LS-DNN architecture. More recently, a comparative study\ \cite{goy2019importance} showed that higher iterates or regularized versions of GS do improve the appearance of the Approximant result $\hat{f}^*$ but do not yield significant improvement in the end output $\hat{f}$ of the DNN. Similar conclusions hold for alternatives to GS, e.g. Gradient Descent. While these alternative schemes are interesting for the LS-DNN method, we chose to not pursue them here. 

The general form of the $(k+1)\text{-th}$ GS iterate from the $k\text{-th}$ iterate is 
\begin{equation}
f[k+1]=\text{arg}\left\{\breathe{3}{6} \mathbf{F}_z^{-1}\left(\breathe{2}{5} 
\sqrt{g}\expb{i \text{ arg}\{\mathbf{F}_z(\ee{if[k]})\}}    \right)    \right\},
\label{GS-iteration}
\end{equation}
where we have taken into account that $\sst{\psi}{inc}=1$. Accordingly, our Approximant is 
\begin{equation}
\hat{f}^*=f[1]=\text{arg}\left\{\breathe{3}{6} \mathbf{F}_z^{-1}\left(\breathe{2}{5} 
\sqrt{g}\expb{i \text{ arg}\{\mathbf{F}_z(\mathbf{1})\}}    \right)    \right\},
\label{eq:GS-1-ite}   
\end{equation}
where $\mathbf{1}$ denotes the function that is uniformly equal to one within the frame \cite{goy2019importance}. 

\begin{figure}[hbt!]
\centering
\includegraphics[width=0.48\textwidth]{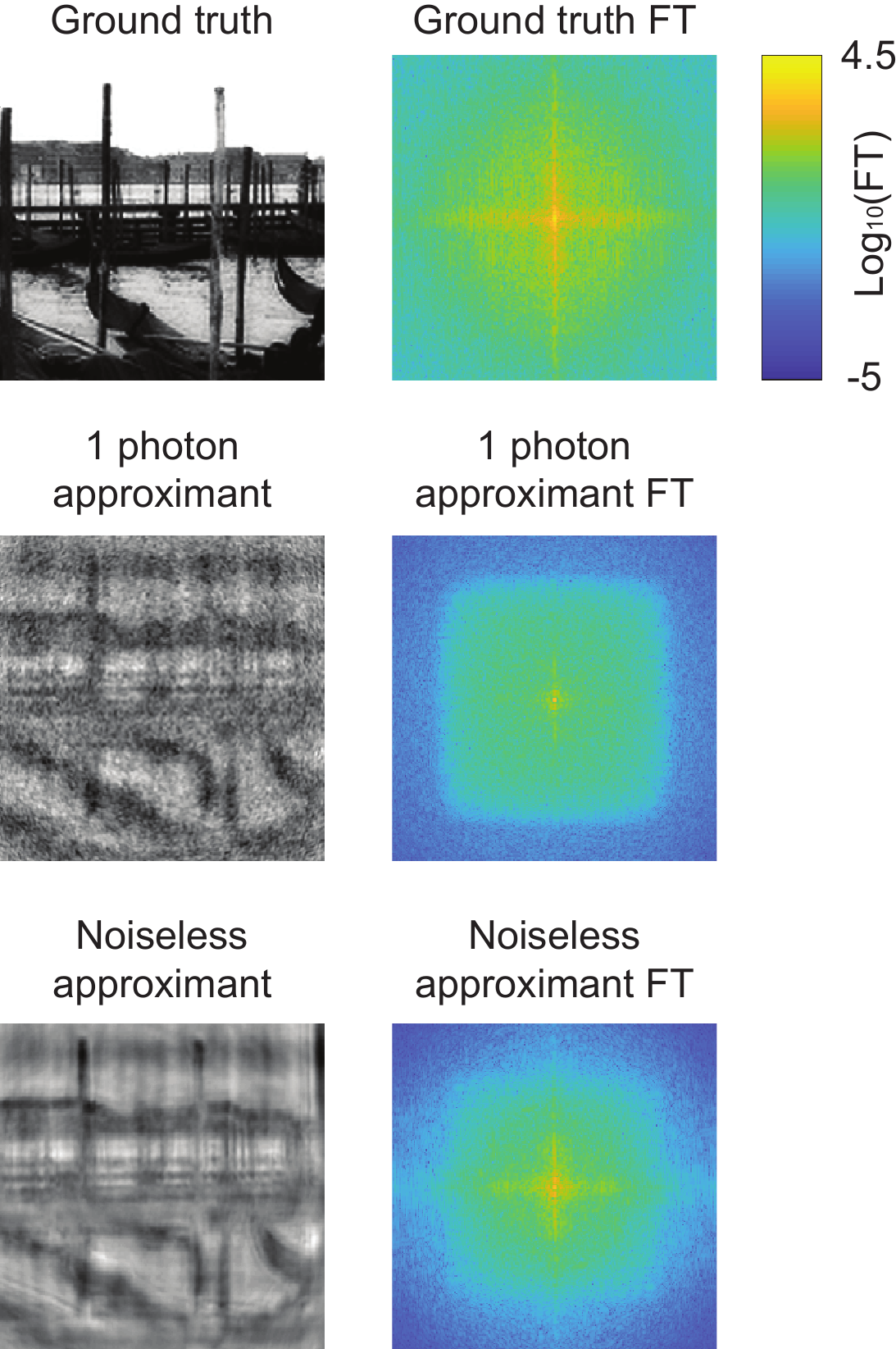} 
\caption{From top to bottom: ground truth image and its 2D Fourier spectrum; noiseless Approximant and its 2D Fourier spectrum; Approximant for 1 photon/pixel illumination and its 2D Fourier spectrum.}
\label{fig: psd-noise-nonoise-compare}
\end{figure}

Figure\ \ref{fig: psd-noise-nonoise-compare} compares the 2D (log-scale) Fourier spectrum magnitude of a ground truth image (from ImageNet \cite{nn:russakovsky15-ImangeNet}), Approximant (\ref{eq:GS-1-ite}) computed without noise, and Approximant (\ref{eq:GS-1-ite}) computed from an input subject to Poisson statistics corresponding to average flux of one photon per pixel. We can see that although the single photon Approximant (which we will later use as the input to the LS-DNN) has a large support in its spectrum, it is the noise that dominates the mid-to-high frequency range. Therefore, the learning process still bears the burden of restoring the correct high frequency contents and relying heavily on high frequency priors, as our DNN-H does, is justified. 

\section{Results}\label{sec:results}

\subsection{Experimental Apparatus and Data Acquisition}\label{subsec:data-acquisition}

\begin{figure*}[hbt!]
\centering
\captionsetup{justification=centering}
\includegraphics[width=\textwidth]{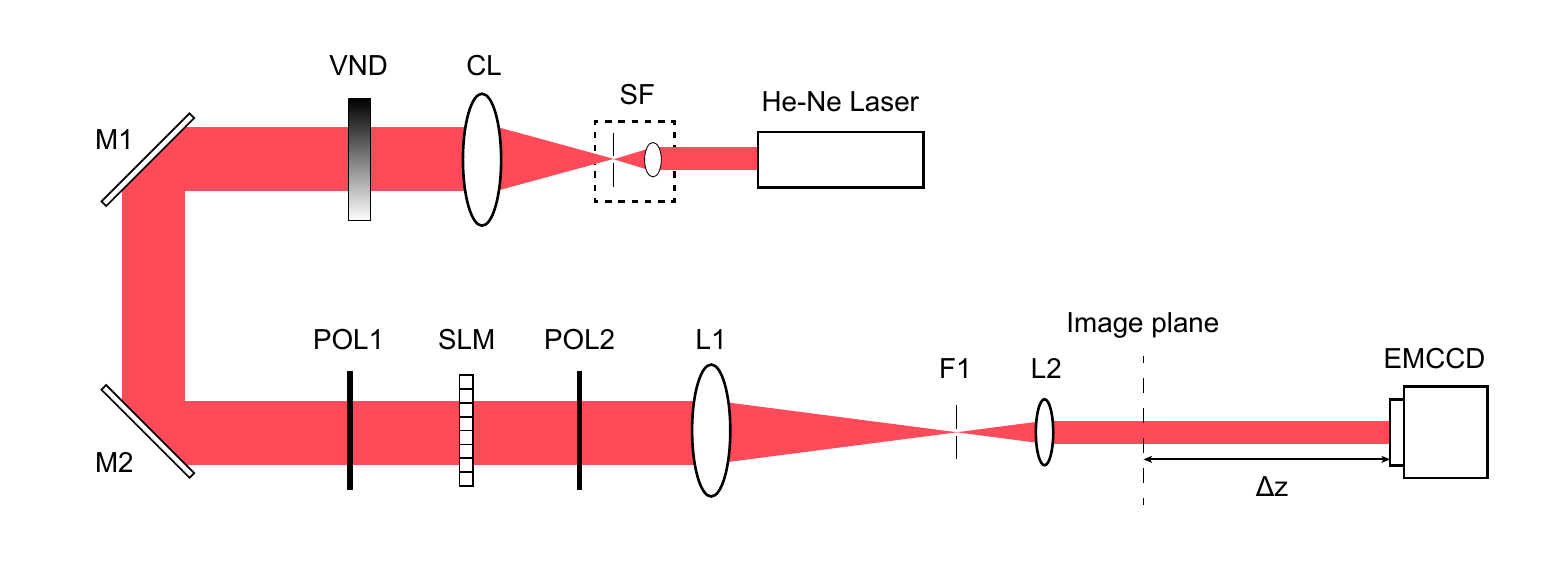}
\caption{Optical apparatus for experiments. SF: spatial filter, CL: collimating lens, VND: variable neutral density filter.}
\label{fig:experimental-set-up}
\end{figure*}

In each experiment carried out to train and test different LS-DNN schemes, 10,000 image objects from ImageNet \cite{nn:russakovsky15-ImangeNet} were successively projected on a phase SLM as phase objects ({\it i.e.}, with phase value at each pixel proportional to the intensity of the corresponding pixel in the original from the database) and their raw images were recorded by the EM-CCD camera at an out-of-focus plane. These 10,000 ground-truth phase images and their corresponding raw intensity images were split into a training set of 9,500 images, a validation set of 450 images and a test set of 50 images. The choice of ImageNet \cite{nn:russakovsky15-ImangeNet} is reasonable, since the low-frequency dominance in its spatial PSD is representative of the broader classes of objects of interest, and therefore, we anticipate our results will generalize well in practical applications. 

The experiments were carried out using the apparatus described in Figure\ \ref{fig:experimental-set-up}. The light source was a continuous wave Helium-Neon gas laser at 632.8nm. The laser beam first passed through a variable neutral density filter (VND) that served the purpose of adjusting the photon flux. The beam was then spatially filtered and expanded into a 18mm diameter collimated pencil and sent onto a transmission spatial light modulator (SLM) with $256\times 256$ pixels, each of size $36\times 36\mu\text{m}$. Phase objects were projected on the SLM and imaged by a telescope (4F system) consisting of lenses L1 (focal length 230mm) and L2 (100mm). The $2.3\times$ reduction factor in the 4F system was designed to reduce the spatial extent of the defocused raw image to approximately fit the size of the camera. An aperture was placed in the Fourier plane to suppress higher diffraction orders due to the periodicity of the SLM pixels. The raw intensity images were captured by a Q-Imaging EM-CCD camera with $1004\times 1002$ square-shaped pixels of size 8$\times$8$\mu$m placed at a distance $z=400\text{mm}$ from the image plane of the 4F system. Additional details about the implementation of the optical apparatus and its numerical simulation with digital Fresnel transforms are in Supplementary Material. 

The photon flux is quantified as the number of photons $p$ received by each pixel on average  for an unmodulated beam, \textit{i.e.} with no phase modulation driving the SLM. During an initial calibration procedure, for different positions of the VND filter, the photon level is measured using a calibrated silicon photodetector placed at the position of the camera. The quoted photon count $p$ is also corrected for the quantum efficiency of the CCD (60\% at $\lambda=632.8\text{nm}$), meaning that we refer to the number of photons actually detected and not the incident number of photons.

Here, we report results for two levels of photon flux $p=9.8\pm 5$\% and $p=1.1\pm 5$\%, respectively quoted in the text as ``10'' and ``1'' photons. The data acquisition, training and testing procedures of the entire LS-DNN architecture were repeated separately for each value of $p$. 

\subsection{Reconstructions and Analysis}\label{subsec:reconstructions}

\begin{figure*}[hbt!]
\centering
\includegraphics[width=\textwidth]{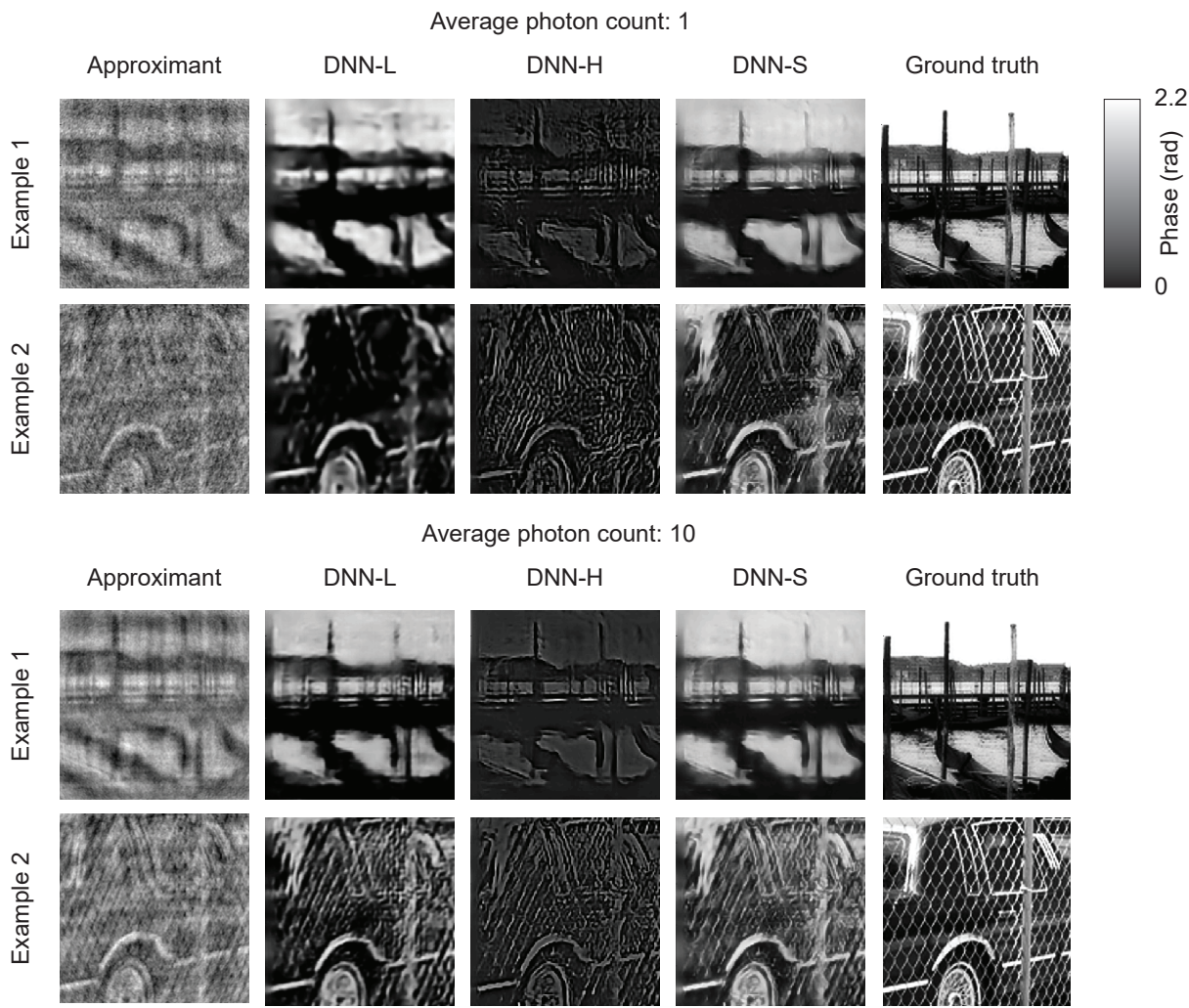}
\caption{Reconstructions by LS-DNN (top: 1 photon/pixel/frame, bottom: 10 photons/pixel/frame); from left to right: Approximant (the input to the LS-DNN system), DNN-L reconstruction \cite{inv:goy2018low}, DNN-H reconstruction ($q=0.5$), DNN-S reconstruction, ground truth.} 
\label{fig:rec}
\end{figure*}

Figure~\ref{fig:rec} shows the reconstructions obtained by the LS-DNN method ($q=0.5$) and its components at fluxes $p = $ 1 photon and 10 photons per pixel, as defined immediately above. As expected, the reconstructions $\hat{f}^{\text{LF}}$ by DNN-L have good fidelity at low spatial frequencies but lose fine details, as in \cite{inv:goy2018low}; whereas the reconstructions $\hat{f}^{\text{HF}}$ by DNN-H look like high-pass filtered versions of the true objects with some additional high-frequency artifacts due to the noise. The reconstructions $\hat{f}$ by DNN-S preserve detail at both low and high frequencies, while significantly attenuating the artifacts. The improvement of $\hat{f}$ over $\hat{f}^{\text{LF}}$ is more pronounced under severe noise, {\it i.e.} in the $p=1$ photon/pixel case. More examples of reconstructions (obtained with $q=0.5$) for the noisier case ($p=1$) are given in the Supplementary Material. 

In Figures~\ref{fig:different-q-PL1} and \ref{fig:different-q-PL10}, we compare reconstructions by LS-DNN with different values of the pre-filtering parameter $q$ for $p=1$ photon and $p=10$ photons per pixel, respectively. The most detail at high frequencies in the DNN-S output is preserved in the range $0.3\lesssim q\lesssim 0.7$. At lower values of $q$, the quality of reconstructions by DNN-S does not noticeably exceed that of DNN-L. This is expected, since in the limit $q=0$, training DNN-H becomes identical to DNN-L. On the other hand, for values $q\gtrsim 0.7$, the DNN-H output is dominated by high-frequency artifacts and again the quality of DNN-S reconstructions regresses to that of DNN-L, since the high-frequency channel is no longer contributing. These observations are valid for both values of $p$, and even stronger for the most severely noise-limited case $p=1$. 

Similar trends are evident according to various quantitative metrics averaged over the entire set of test examples compared to the true phase signals $f$, summarized in Table~\ref{table:all}. For comparisons we used the PCC, defined as in (\ref{eq:npcc}) but without the minus sign; peak signal-to-noise ratio (PSNR)\ \cite{gupta2011modified}; and structural similarity index metric (SSIM)\ \cite{wang2003multiscale,inv:wang04-SSIM}. As we noted in Section \ref{sec:methods}.\ref{sub:LS}, DNN-H is trained with spectrally pre-filtered version of the true object $f$ so a quantitative comparison of its output with the ground truth does not make sense.  

It is noteworthy that in both visualization and quantitative comparisons of Figures~Figures~\ref{fig:different-q-PL1}-\ref{fig:different-q-PL10} and Table~\ref{table:all}, respectively, the performance of DNN-S remains approximately the same within the range $0.3\lesssim q\lesssim 0.7$. This is desirable as it suggests that one need not prefilter exactly with the inverse of the PSD power law. This further suggests that for datasets that do not represent natural images and may obey power laws different than (\ref{eq:psd}), not knowing the exact value of $q$ may not be catastrophic. We have not tested this hypothesis exhaustively, as it is beyond the scope of this paper. 

In Table \ref{table:compare DNN-L-3} and in Supplementary Material, we also analyze the case of a bigger DNN (denoted DNN-L-3) with computational capacity equal to the sum of DNN-L, H and S together, but trained with un-filtered examples, and show that DNN-L-3 cannot achieve reconstructions of equal quality. Therefore, the improvements over \cite{inv:goy2018low} resulted from the training procedure followed in the LS-DNN method and not simply by brute force due to the use of larger computational capacity. 

\begin{figure*}[hbt!]
    \centering
    \captionsetup{justification=centering}
    \includegraphics[width=\textwidth]{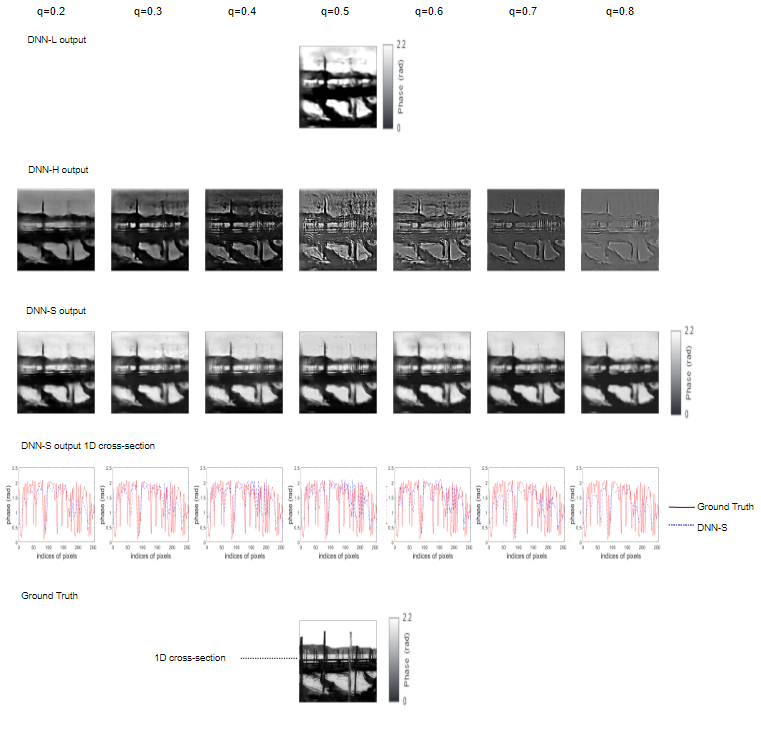}
    \caption{Comparisons of LS-DNN reconstructions under different $q's$ for $p=1$ photon/ pixel. Rows from top to bottom: DNN-L output; DNN-H output under different $q$'s; DNN-S under different $q$'s;  1D cross-section (along the dashed line in the row below) of DNN-S output under different $q's$; ground truth.}
    \label{fig:different-q-PL1}
\end{figure*}

\begin{figure*}[hbt!]
    \centering
    \captionsetup{justification=centering}
    \includegraphics[width=\textwidth]{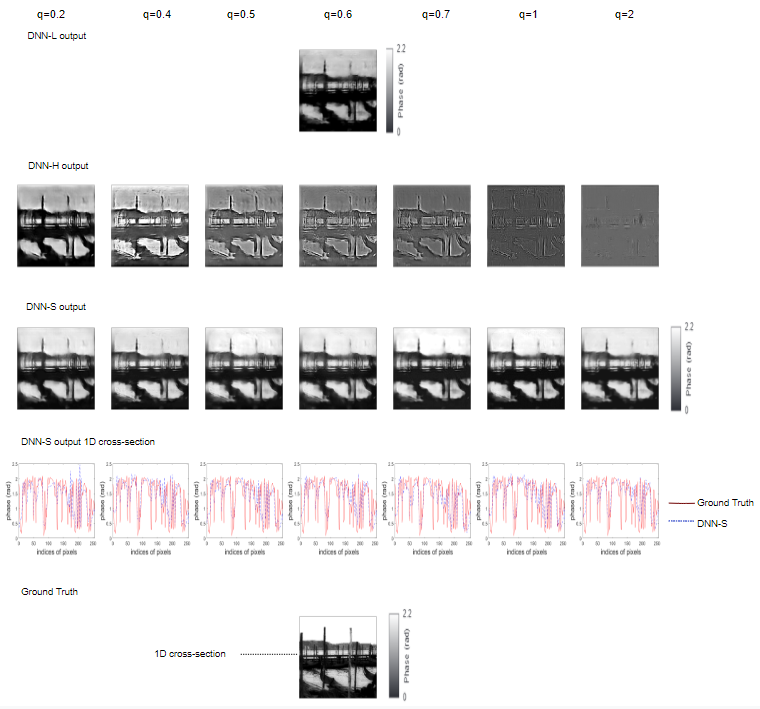}
    \caption{Comparisons of LS-DNN reconstructions under different $q's$ for $p=10$ photon/ pixel. Rows from top to bottom: DNN-L output; DNN-H output under different $q$'s; DNN-S under different $q$'s; 1D cross-section (along the dashed line in the row below) of DNN-S output under different $q's$; ground truth.}
    \label{fig:different-q-PL10}
\end{figure*}

\begin{table*}[hbt!]
\begin{tabular}{|l||c|c||c|c||c|c|}
 \hline
& \multicolumn{2}{c||}{Average PCC $\pm$ std.dev}  & \multicolumn{2}{c||}{Average PSNR $\pm$ std.dev (dB)} & \multicolumn{2}{c|} {Average SSIM $\pm$ std.dev}  \\
 \hline
  & $p=1$ & $p=10$ & $p=1$ & $p=10$ & $p=1$ & $p=10$ \\
 \hline
 Approximant $\hat{f}^*$ & 0.149$\pm$ 0.067 & 0.181 $\pm$ 0.082 & 8.082 $\pm$ 3.979 & 8.098 $\pm$ 3.986 & 0.220 $\pm$ 0.105 & 0.222 $\pm$ 0.106 \\
 DNN-L output $\hat{f}^{\text{LF}}$ & 0.808 $\pm$ 0.099 & 0.880 $\pm$ 0.059 & 16.207 $\pm$ 2.466 & 18.299 $\pm$ 2.536 & 0.875 $\pm$ 0.071  & 0.925 $\pm$ 0.039 \\
 \hline 
 DNN-S output $\hat{f}$ ($q=0.1$) & 0.848 $\pm$ 0.075 & 0.892$\pm$ 0.053 & 17.511 $\pm$ 2.317 & 18.426 $\pm$ 2.321 & 0.905 $\pm$ 0.054 & 0.929 $\pm$ 0.048\\
 DNN-S output $\hat{f}$ ($q=0.2$)  & 0.858$\pm$ 0.074 & 0.898 $\pm$ 0.050 & 17.758$\pm$ 2.523 & 18.469 $\pm$ 2.469 & 0.903 $\pm$ 0.056 & 0.926 $\pm$ 0.036\\
 DNN-S output $\hat{f}$ ($q=0.3$) & 0.859 $\pm$ 0.071 & 0.896 $\pm$ 0.050 & 17.769 $\pm$ 2.420 & 18.487 $\pm$ 2.313 & 0.912 $\pm$ 0.048 & 0.930 $\pm$ 0.043\\
 \hline
 DNN-S output $\hat{f}$ ($q=0.4$) &  0.865 $\pm$ 0.067 & 0.897 $\pm$ 0.050 & 18.116 $\pm$ 2.319 & 18.634$\pm$ 2.380 & 0.915 $\pm$ 0.061 & 0.934 $\pm$ 0.034\\
 DNN-S output $\hat{f}$ ($q=0.5$) & 0.866 $\pm$ 0.069 & 0.899 $\pm$ 0.048 & 17.940 $\pm$ 2.281 & 18.902 $\pm$ 2.443 & 0.915 $\pm$ 0.051 & 0.937 $\pm$ 0.034 \\
 DNN-S output $\hat{f}$ ($q=0.6$) & 0.860 $\pm$ 0.071 & 0.898 $\pm$ 0.050 & 17.880 $\pm$ 2.455 & 18.940 $\pm$ 2.320 & 0.915 $\pm$ 0.062 & 0.935 $\pm$ 0.038 \\
 \hline
 DNN-S output $\hat{f}$ ($q=0.7$) & 0.859$\pm$ 0.088 & 0.892 $\pm$  0.051 & 17.827 $\pm$ 2.140 & 18.487 $\pm$ 2.311 & 0.911 $\pm$ 0.070 & 0.927 $\pm$ 0.047 \\ 
 DNN-S output $\hat{f}$ ($q=0.8$) & 0.845$\pm$ 0.081 & 0.887 $\pm$  0.050 & 17.375 $\pm$ 2.221 & 18.445 $\pm$ 2.470 & 0.902 $\pm$ 0.059 & 0.934 $\pm$ 0.041\\ 
 DNN-S output $\hat{f}$ ($q=1$) & 0.825 $\pm$ 0.111 & 0.883 $\pm$ 0.052 & 16.991$\pm$  2.008 & 18.484$\pm$ 2.363 & 0.894 $\pm$ 0.075& 0.934 $\pm$ 0.042\\
 DNN-S output $\hat{f}$ ($q=2$) & 0.819 $\pm$ 0.077 & 0.882 $\pm$ 0.055 &16.820 $\pm$ 2.254 & 18.391 $\pm$ 2.600 & 0.893 $\pm$ 0.058 & 0.933 $\pm$ 0.037\\
 \hline 
\end{tabular}
\caption{Quantitative comparison of reconstructions by Approximant, DNN-L and DNN-S. Each entry takes the form of 'average $\pm$ standard deviation'.}
\label{table:all}
\end{table*}

\begin{table*}[hbt!]
\begin{tabular}{|l||c|c||c|c||c|c|}
 \hline
& \multicolumn{2}{c||}{Average PCC $\pm$ std.dev}  & \multicolumn{2}{c||}{Average PSNR $\pm$ std.dev (dB)} & \multicolumn{2}{c|} {Average SSIM $\pm$ std.dev}  \\
 \hline
  & $p=1$ & $p=10$ & $p=1$ & $p=10$ & $p=1$ & $p=10$ \\
 \hline
 Approximant $\hat{f}^*$ & 0.149$\pm$ 0.067 & 0.181 $\pm$ 0.082 & 8.082 $\pm$ 3.979 & 8.098 $\pm$ 3.986 & 0.220 $\pm$ 0.105 & 0.222 $\pm$ 0.106 \\
 DNN-L output $\hat{f}^{\text{LF}}$ & 0.808 $\pm$ 0.099 & 0.880 $\pm$ 0.059 & 16.207 $\pm$ 2.466 & 18.299 $\pm$ 2.536 & 0.875 $\pm$ 0.071  & 0.925 $\pm$ 0.039 \\
 DNN-L-3 output $\hat{f}^{\text{L-3}}$ & 0.811 $\pm$ 0.125 & 0.880 $\pm$ 0.117 & 16.209 $\pm$ 2.522 & 18.268 $\pm$ 2.030 & 0.835 $\pm$ 0.092 & 0.926 $\pm $ 0.073\\
 DNN-S output $\hat{f}$ ($q=0.5$) & 0.866 $\pm$ 0.069 & 0.899 $\pm$ 0.048 & 17.940 $\pm$ 2.281 & 18.902 $\pm$ 2.443 & 0.915 $\pm$ 0.051 & 0.937 $\pm$ 0.034 \\

 \hline
\end{tabular}
\caption{Quantitative comparison of reconstructions by Approximant, DNN-L-3 and DNN-S (for $q=0.5$). Each entry takes the form of 'average $\pm$ standard deviation'.}
\label{table:compare DNN-L-3}
\end{table*}

\begin{figure*}[hbt!]
    \centering
    \includegraphics[width=\textwidth]{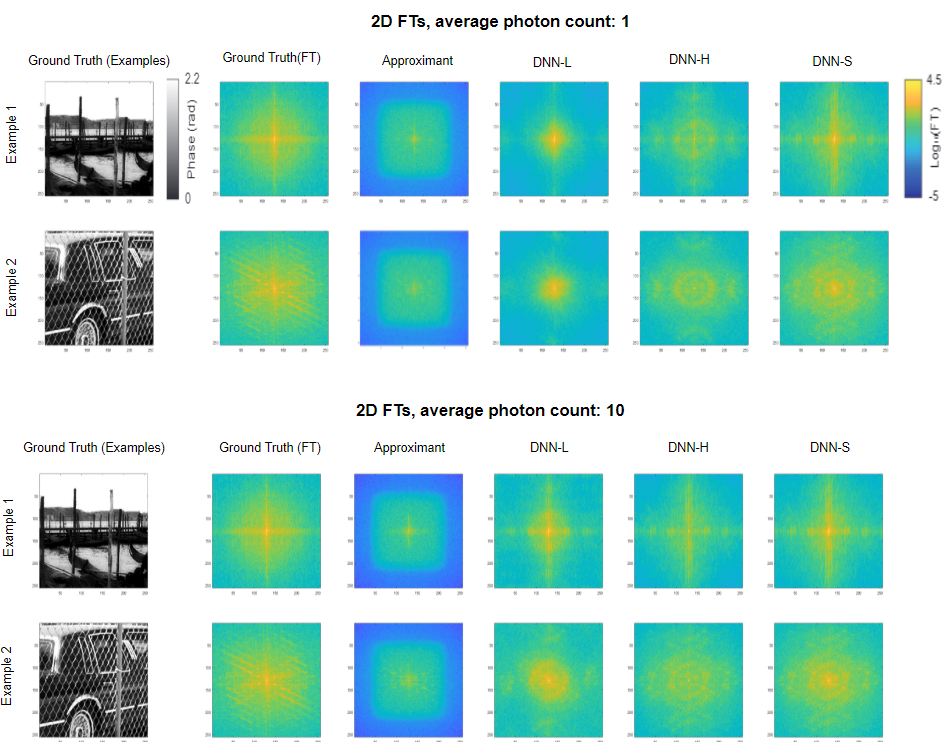}
    \caption{Fourier spectra of two test examples and their reconstructions from the components of the LS scheme. }
    \label{fig:PSD_2D}
\end{figure*}

To further study the behavior of the LS components in the low and high spatial frequency bands, we studied the reconstructions in the Fourier domain. Figure~\ref{fig:PSD_2D} shows the spectra (2D Fourier transforms) of two randomly selected test examples. Figure \ref{fig:PSD_1D} and Figure~5 in Supplementary Material show normalized diagonal cross-sections of the PSD averaged over all the 50 test images, for $p=1$ and $10$ photons per pixel, respectively. These plots illustrate that DNN-L and DNN-H's outputs are depleted at the high and low frequencies, respectively, with the losses being more severe in the noisy case $p=1$; whereas DNN-S's output mostly recovers the frequency content at both bands, albeit still with some minor loss at high frequencies.

\begin{figure*}[hbt!]
\centering
\includegraphics[width=0.8\textwidth]{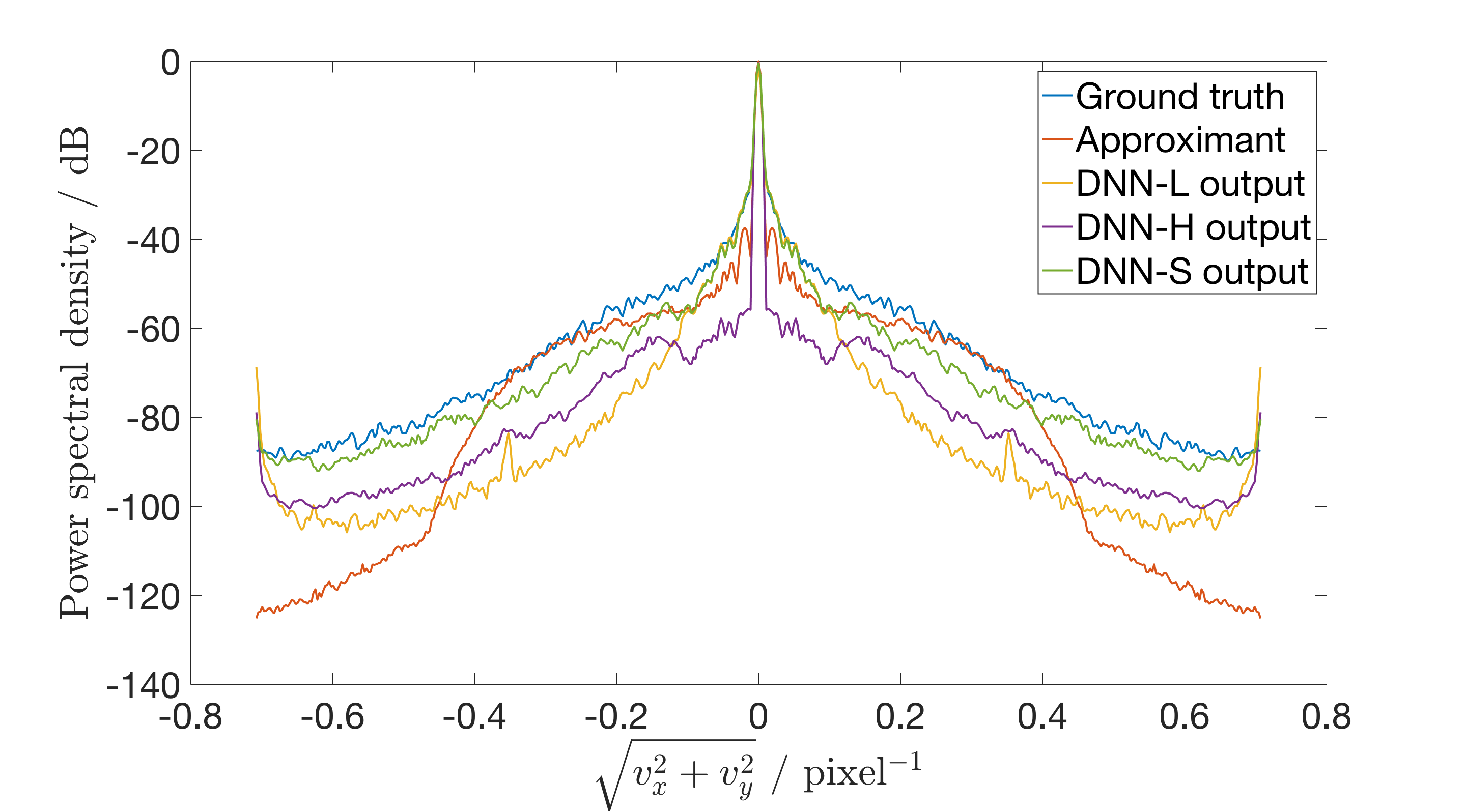}
\caption{1D diagonal cross-sections of the average 2D power spectral density (PSD) of 50 test images for $p=1$ photon per pixel. The case of $p=10$ photons per pixel is in the Supplementary Material.} 
\label{fig:PSD_1D}
\end{figure*}

Lastly, we experimentally characterized the spatial resolution of the LS-DNN reconstructions, {\it i.e.} the ability of DNN-S to resolve two pixels at nearby locations having phase delay higher than the rest of the signal. Similar analyses were carried out in \cite{inv:IDiffNet,inv:PhENN-spectral-premod}, where the methodology was also described in detail. In the work presented here, we carried out the analysis under ample illumination, {\it i.e.} not under strong Poisson statistics. We made that choice because spatial resolution under highly noisy conditions becomes non-trivially coupled to the noise statistics, and a complete investigation would have been outside the scope of the present investigation. The results are shown in section {\color{blue}6} in Supplementary Material.

\section{Concluding Remarks} \label{sec:conclusions}
The LS-DNN reconstruction scheme\ \cite{deng2018learning} for quantitative phase retrieval has been shown to be resilient to highly noisy raw intensity inputs while preserving high spatial frequency detail better than\ \cite{inv:goy2018low}. The reconstructions are also quite robust to variations of the pre-filtering power law $q$ around the value $\approx 1/2$ following from natural image statistics. Beyond the scope of the work reported here, further improvements may be obtained through modifying the architecture of the DNNs used to process and recombine the two spatial frequency bands. It is also possible that for highly restricted datasets, $q$ needs to be investigated further. However, our observations for natural images suggests that the LS-DNN approach is relatively insensitive to $q$ under fairly wide conditions.

Another obvious alternative strategy is to split the signals into more than two bands, then process and recombine these multiple bands with a synthesizer DNN, according to the LS scheme. While we did not investigate this approach in detail here, we expect it to present a trade-off between the improvements and the complexity of having to train multiple neural networks implying the need for more examples and the danger of poor generalization. \\

\section*{Acknowledgments}

This work was supported by the Intelligence Advanced Research Projects Activity (IARPA) grant No. FA8650-17-C-9113 and by the SenseTime company. I. Kang was supported in part by KFAS (Korea Foundation for Advanced Studies) scholarship. We are grateful to Kwabena Arthur for useful discussions and critique of the manuscript.

\bibliography{main}

\end{document}